\documentclass[symmetry,article,accept,moreauthors]{Definitions/mdpi} 

\firstpage{1} 
\pubvolume{1}
\issuenum{1}
\articlenumber{0}
\pubyear{2024}
\copyrightyear{2024}
 \externaleditor{{Academic Editor: Firstname Lastname}}
\datereceived{3 March 2024} 
\daterevised{18 March 2024} 
\dateaccepted{25 March 2024} 
\datepublished{ } 
\hreflink{{http://doi.org/10.3390/sym16040394}} 

\Title{Symmetry Transformations in Cosmological and Black Hole Analytical Solutions}

\TitleCitation{{Symmetry Transformations in Cosmological and Black Hole Analytical Solutions}}

\Author{Edgar A. Le\'{o}n *
$^{,\dagger}$\orcidA{}  and Andr\'{e}s Sandoval-Rodr\'{i}guez~$^\dagger$\orcidB{}}

\AuthorNames{Edgar A. Le\'{o}n  and Andr\'{e}s Sandoval-Rodr\'{i}guez}

\AuthorCitation{Le\'{o}n, E.A.; Sandoval-Rodr\'{i}guez, A}

\address[1]{%
 Facultad de Ciencias F\'{\i}sico-Matem\'{a}ticas, Universidad Aut\'{o}noma de Sinaloa, 80010 Culiac\'{a}n, Mexico
;  andres.fcfm@uas.edu.mx}
\corres{\hangafter=1 \hangindent=1.05em \hspace{-0.82em}Correspondence: ealeon@uas.edu.mx}

\firstnote{\hangafter=1 \hangindent=1.05em \hspace{-0.82em}These authors contributed equally to this work.}

\abstract{We analyze the transformation of a very broad class of metrics that can be expressed 
 in terms of static coordinates. Starting from a general ansatz, we obtain a relation for the parameters in which one can impose further symmetries or restrictions. One of the simplest restrictions leads to FLRW cases, while transforming from the initial static to other static-type coordinates can lead to near horizon coordinates, Wheeler--Regge, and isotropic coordinates, among others. As less restrictive cases, we show an indirect route for obtaining Kruskal--Szekeres within this approach, as well as Lema\^{\i}tre coordinates. We use Schwarzschild spacetime as a prototype for testing the procedure in individual cases. However, application to other
spacetimes, such as de-Sitter, Reissner--Nordstr\"{o}m, and Schwarzschild de Sitter, can be readily generalized.}


\begin{document}

\section{Introduction}

The role of coordinates in general relativity is a subtle issue
which must be taken with caution for each one of the individual solutions to
Einstein's field equations. There is a class of coordinate systems that are
useful in a specific patch of the spacetime. Usually, these types of
transformation are suitable for very specific physical interpretations,
such as when one can attach them to a type of observers. This occurs, for
instance, in Painlev\'{e}--Gullstrand (or rain) coordinates for black hole
static solutions~\cite{ref-book1}. Isotropic coordinates for static spacetimes
can also be included in this class, which allow for comparing asymptotic
behavior at spatial infinity with a Newtonian metric~\cite{ref-book2}.

A second type of transformations includes those that allow for the continuity of
spacetime to be considered. In particular, the new coordinates can clarify
the coordinate character of singularities associated with cosmological and
event horizons. For example, the Lema\^{\i}tre system provided one of the first
descriptions for the Schwarzschild case, which allowed for a continuous crossing
of the event horizon. Eddington--Finkelstein coordinates, adapted to null
coordinates, also satisfy this property~\cite{ref-journal1, ref-book3}.

Other coordinates are useful to visualize relevant regions of the spacetime
in the same diagram or even expand it to include possible copies of it. In
fact, we can interpret the Schwarzschild and other spherically symmetric
spacetimes as a section of the broader Kruskal--Szekeres spacetime~\cite{ref-journal2,ref-journal3,ref-book4,ref-book5}. Finally, other representations allow us to
visualize effects such as how space is curved (e.g., with embedding diagrams)
or to visualize the behavior of the complete spacetime in a compact diagram,
as is the case with conformal diagrams~\cite{ref-journal4,ref-journal5}. It is worth emphasizing that some coordinates may possess properties that make
them ubiquitous in this classification.

In this article, we analyze transformations for spherically symmetric
spacetimes that can be described with static coordinates. Some of the most
paradigmatic solutions in cosmology and black hole theory can be formulated
in these terms. Some examples include the following: Schwarzschild, Reissner--N\"{o}%
rdstrom, Schwarzschild--de Sitter, de Sitter, and anti-de Sitter spacetimes,
just to name a few. We start precisely with the static version, where $%
g_{11}g_{00}=-1$~\cite{ref-journal6}. From there, we consider the transformation
to a generic form that maintains emphasis on the isotropy with respect to $%
r=0$. We obtain a crucial equation, from which we can impose restrictions
that allow us to solve for specific cases. Although we maintain the spirit
of showing general relationships, we mostly use the Schwarzschild solution as
a specific example of the procedures. Our treatment can encompass a wide class of metric solutions and, as we shall see, there is also a direct relationship with respect to the ADM formulation and the identification of underlying symmetries of the spacetimes~involved.

The structure for the rest of this article is as follows. We present in
Section~\ref{sec2} the initial form of the metric and also derive useful relations that
lead to the master equation that dictates the way to obtain specific
solutions. Section \ref{sec3} is dedicated to briefly discussing how and which FLRW
solutions emerge. Furthermore, starting from the restricted Friedmann
equations that appear, we summarize the cosmological solutions that can be
put into static form. Section \ref{sec4} has a broader scope, as it includes many
possibilities: those leading from static coordinates to other static
coordinates. Specifically, we obtain near-horizon coordinates,
Regge--Wheeler, isotropic coordinates, and a transformation that resembles Lema%
\^{\i}tre coordinates. In Section~\ref{sec5}, we obtain, Kruskal--Szekeres as well as the
correct Lema\^{\i}tre coordinates for Schwarzschild spacetime. Both are
examples of specific coordinates with the characteristics analyzed
throughout the article, but without the restrictions of the last previous
two sections. Finally, in Section~\ref{sec6}, we make some remarks about the method and
the~results.

\section{The General Transformations}\label{sec2}

It is well known that Einstein equations,%
\begin{equation}
R_{\mu \nu }-\frac{1}{2}g_{\mu \nu }(R-2\Lambda )=8\pi GT_{\mu \nu }, 
\tag{1}
\end{equation}%
admit several analytic solutions that can be put in the following form:%
\begin{equation}
ds_{(1)}^{2}=g_{\mu \nu }dx^{\mu \nu }=-fdt^{2}+f^{-1}dr^{2}+r^{2}d\Omega
^{2}.  \tag{2}
\end{equation}%
Here, we have $f=f(r)$ and $d\Omega ^{2}=d\theta ^{2}+\sin ^{2}\theta $ $%
d\phi ^{2}$. This encompasses a broad class of spherically symmetric metrics
that can be put in a static form~\cite{ref-journal7,ref-journal8,ref-journal9}. There is a subclass of FLRW-metrics can be put in the form of (2), such as the de Sitter (dS) and AdS spaces, as well as Milne and Lanczos universes. Other classes of possibilities include black hole solutions, such as Schwarzschild, Reissner--Nordstr\"{o}m, and Schwarzschild--de Sitter, to name a few.

From the coordinates that define the metric $ds_{(1)}^{2}$, namely, $%
x^{\alpha }=(t,r,\theta , \phi )$, we consider a very general
transformation to a metric defined in the new coordinates $x^{\alpha
^{\prime }}=(\tau ,\rho ,\theta , \phi )$:%
\begin{equation}
ds_{(2)}^{2}=\gamma _{\alpha ^{\prime }\beta ^{\prime }}dx^{\alpha ^{\prime
}}dx^{\beta ^{\prime }}=-N^{2}d\tau ^{2}+b^{2}\left( g^{2}d\rho ^{2}+\rho
^{2}d\Omega ^{2}\right) .  \tag{3}
\end{equation}%
The metric components have dependence $N=N(\tau ,\rho )$, $b=b(\tau ,\rho )$
and $g=g(\rho )$. Notice that $N$ corresponds to the lapse function in
the ADM decomposition, given by%
\begin{equation}
ds^{2}=-N^{2}d\tau ^{2}+h_{ij}(N^{i}d\tau +dx^{i})(N^{j}d\tau +dx^{j}), 
\tag{4}
\end{equation}%
where the Latin indices run in the spatial sections from $1$ to $3$~\cite{ref-journal10,ref-journal11}. Comparing these last relations, we have $N_{i}=h_{ij}N^{j}=0$ (spatial slices orthogonal to time direction). Clearly, $h_{ij}=\gamma _{i^{\prime }j^{\prime }}$ is also diagonal. Even with this identification, it is useful to keep in mind that, in general, one cannot make the association $fdt^{2}=N^{2}d\tau ^{2}$.

Furthermore, take into account the extrinsic curvature, defined by%
\begin{equation*}
K_{ij}=\tilde{\nabla}_{i}N_{j}+\tilde{\nabla}_{j}N_{i}-\frac{1}{2N}\frac{%
\partial h_{ij}}{\partial \tau }=-\frac{1}{N}\frac{\dot{b}}{b}h_{ij}.   \tag{5}
\end{equation*}

Since the scalar curvature $R$ involves (except to a total derivative) $K_{ij}$ and its norm $K=h^{ij}K_{ij}$, if both are zero, then $R$ is just the tridimensional scalar curvature. In terms of this, we have that the simplest nontrivial class of solutions involving this quantity results when $b=b(\tau) $.

All this suggests that we should include FLRW solutions as a nontrivial and direct case study. For instance, $N=1$ with $bg=\exp (\sqrt{\Lambda /3}\tau )$ corresponds to a de Sitter space, which is the solution of main importance as the asymptotic limit for the standard $\Lambda $-CDM cosmology. In general, the gauge where the lapse function is $N=const.$  $N=1$, with $b=b(\tau )$, is the adequate ansatz for FLRW metrics. Other restrictions can lead to Kruskal or other types of coordinates representing the Schwarzschild spacetime, and so on~\cite{ref-journal12}.

Now, the general metric transformation that interest us is

\begin{equation}
g_{\alpha ^{\prime }\beta ^{\prime }}=\frac{\partial x^{\alpha }}{\partial
x^{\alpha ^{\prime }}}\frac{\partial x^{\beta }}{\partial x^{\beta ^{\prime
}}}g_{\alpha \beta },  \tag{6}
\end{equation}%
where the indices run from $0$ to $3$ in the order mentioned for both sets of
coordinates. From here on, an overdot denotes a partial derivative with respect to $%
\tau $, such as in $\dot{t}=\partial t/\partial \tau $. In a similar way, a
prime denotes a partial derivative with respect to $\rho $, such as in $%
r^{\prime }=\partial r/\partial \rho $. The angular components of (6), $%
g_{2^{\prime }2^{\prime }}$ and $g_{3^{\prime }3^{\prime }}$, are tantamount
to $r=b\rho $. This implies that $\dot{r}=\dot{b}\rho $ and $r^{\prime
}=b+b^{\prime }\rho $. On the other hand, the expansion for $g_{0^{\prime
}0^{\prime }}$, $g_{1^{\prime }1^{\prime }}$, and $g_{0^{\prime }1^{\prime }}$
in (6) leads, after rearrangements, to%
\begin{equation}
\dot{t}^{2}=f^{-2}(fN^{2}+\dot{b}^{2}\rho ^{2}),  \tag{7}
\end{equation}%
\begin{equation}
t^{\prime 2}=f^{-2}\left[ (b+b^{\prime }\rho )^{2}-fb^{2}g^{2}\right] 
\tag{8}
\end{equation}%
and%
\begin{equation}
f^{2}\dot{t}t^{\prime }=\dot{b}\rho (b+b^{\prime }\rho ),  \tag{9}
\end{equation}%
respectively. By inserting (7) and (8) in (9), we obtain, after simplification,
the useful relation%
\begin{equation}
f=b^{-2}g^{-2}(b+b^{\prime }\rho )^{2}-N^{-2}\dot{b}^{2}\rho ^{2}.  \tag{10}
\end{equation}%
In fact, by inserting it back in (7) and (8), they simplify to%
\begin{equation}
\dot{t}=Nf^{-1}b^{-1}g^{-1}(b+b^{\prime }\rho )  \tag{11}
\end{equation}%
and%
\begin{equation}
t^{\prime }=N^{-1}f^{-1}b\dot{b}g\rho .  \tag{12}
\end{equation}

The partial derivation of (11) with respect to $\rho $ leads to%
\begin{equation}
\begin{array}{c}
\frac{\partial ^{2}t}{\partial \rho \partial \tau }=f^{-2}b^{-2}g^{-2}%
\{Nfbg(2b^{\prime }+b^{\prime \prime }\rho ) \\ 
\\ 
+[N^{\prime }fbg-N\left( bgf^{\prime }+fgb^{\prime }+fbg^{\prime }\right)
](b+b^{\prime }\rho )\}.%
\end{array}
\tag{13}
\end{equation}%
In a similar way, from (12), we have%
\begin{equation}
\frac{\partial ^{2}t}{\partial \tau \partial \rho }=N^{-2}f^{-2}g\rho
\lbrack Nf(\dot{b}^{2}+b\ddot{b})-(N\dot{f}+f\dot{N})b\dot{b}].  \tag{14}
\end{equation}%
We equate these two relations. It is useful to note that from (10), one
explicitly obtains%
\begin{equation}
\dot{f}=2b^{-3}g^{-2}\rho (b+b^{\prime }\rho )(b\dot{b}^{\prime }-\dot{b}%
b^{\prime })+2N^{-3}(\dot{N}\dot{b}^{2}-N\dot{b}\ddot{b})\rho ^{2}.  \tag{15}
\end{equation}%
and %
\begin{equation}
\begin{array}{c}
f^{\prime }=2b^{-3}g^{-3}(b+b^{\prime }\rho )\left[ bg(b^{\prime }+b^{\prime
\prime }\rho )-b^{\prime 2}g\rho -bg^{\prime }(b+b^{\prime }\rho )\right] \\ 
\\ 
+2N^{-3}\dot{b}\rho \left[ N^{\prime }\dot{b}\rho -N(\dot{b}^{\prime }\rho +%
\dot{b})\right] .%
\end{array}
\tag{16}
\end{equation}%
Using these two relations in the equality between (13) and (14), the following relation appears after some algebraic steps:%
\begin{equation}
\begin{array}{c}
N^{4}\left( 1+\frac{b^{\prime }}{b}\rho \right) ^{2}\left[ \left( \frac{%
g^{\prime }}{g}+\frac{N^{\prime }}{N}\right) \left( 1+\frac{b^{\prime }}{b}%
\rho \right) +\frac{b^{\prime 2}}{b^{2}}\rho -\left( \frac{b^{\prime }}{b}+%
\frac{b^{\prime \prime }}{b}\rho \right) \right] \\ 
+N^{2}g^{2}\rho ^{2}\left \{ \left[ \left( \frac{g^{\prime }}{g}+\frac{%
N^{\prime }}{N}-\frac{b^{\prime }}{b}\right) +4\frac{\dot{b}^{\prime }}{\dot{%
b}}\right] \left( 1+\frac{b^{\prime }}{b}\rho \right) -\frac{b^{\prime
\prime }}{b}\rho \right \} \dot{b}^{2} \\ 
-N^{2}b^{2}g^{2}\rho \left( 1+\frac{b^{\prime }}{b}\rho \right) ^{2}\left( 
\frac{\ddot{b}}{\dot{b}}+\frac{\dot{b}}{b}-\frac{\dot{N}}{N}\right) \frac{%
\dot{b}}{b}+2N^{2}b^{2}g^{2}\rho \frac{\dot{b}^{2}}{b^{2}} \\ 
=b^{4}g^{4}\rho ^{3}\left( \frac{\ddot{b}}{\dot{b}}-\frac{\dot{b}}{b}-\frac{%
\dot{N}}{N}\right) \frac{\dot{b}^{3}}{b^{3}}.%
\end{array}
\tag{17}
\end{equation}%
This is a crucial equation for the main part of the rest of this article,
where we restrict it to several cases of interest, remembering that the generic
form of the metric is%
\begin{equation}
ds_{(2)}^{2}=-N^{2}d\tau ^{2}+b^{2}\left( g^{2}d\rho ^{2}+\rho ^{2}d\Omega
^{2}\right) .  \tag{18}
\end{equation}%
From here on, we assume some restrictions in the variables $N,$ $b$,
and $g$ to see what type of space is obtained, and at the same time, to
visualize if the identification turns out to be unique.

\section{FLRW Cases}\label{sec3}

When $N=1$ and $b=b(\tau )$ in this metric, one has, precisely, the ansatz for
the co-moving coordinates $(\rho ,\theta ,\phi )$ of an isotropic and
homogeneous expanding universe. A substitution of $N=1$ and $b^{\prime }=0$ in
(17) yields massive simplifications:%
\begin{equation}
\frac{g^{\prime }}{g^{3}\rho }=b\ddot{b}-\dot{b}^{2}.  \tag{19}
\end{equation}%
Since the dependences of the left and right parts of the equality are on $%
\rho $ and $\tau $, respectively, this relation is equal to a constant $%
\kappa $. This allows us to integrate $g^{\prime }g^{-3}=\kappa \rho $,
yielding the important relation%
\begin{equation}
g^{2}=\frac{1}{B-\kappa \rho ^{2}},  \tag{20}
\end{equation}%
where $B$ is an integration constant and $\kappa $ can have any sign. Now,
since locally---where the term $\kappa \rho ^{2}$ can be neglected---the
homogeneous space is flat, $B=1$. That is, we have recovered the
usual FLRW solution.

The right hand of Equation~(19) is also equal to $\kappa $, allowing us to write
the relation as%
\begin{equation}
\frac{\ddot{b}}{b}=\frac{\dot{b}^{2}+\kappa }{b^{2}}.  \tag{21}
\end{equation}%
This equation can be integrated, for instance, by making the reduction of
order $w(\tau )=\dot{b}^{2}+\kappa $, and then $\dot{w}=2\dot{b}\ddot{b}$,
turning (21) into%
\begin{equation}
\frac{\dot{w}}{w}=2\frac{\dot{b}}{b},  \tag{22}
\end{equation}%
with solution $w=\Gamma b^{2}$, where $\Gamma $ is another integration
constant. Then, we can reinterpret (21) as two relations: 
\begin{equation}
\frac{\dot{b}^{2}+\kappa }{b^{2}}=\Gamma ,  \tag{23}
\end{equation}%
and 
\begin{equation}
\frac{\ddot{b}}{b}=\Gamma .  \tag{24}
\end{equation}%
These two relations can be recognized as the two Friedmann equations, with cosmological constant $\Lambda =3\Gamma $ as the only source of the energy momentum tensor~\cite{ref-journal13}.

Even more, we have argued before that (20) should be written as $g^{-2}=1-\kappa \rho ^{2}$. The substitution of this, as well as $r=b\rho$ and the relation $\dot{b}^{2}=\Gamma b^{2}-\kappa $ (see Equation~(23)), in (10) leads to $f=1-\Gamma r^{2}$.

Depending on the value of $\Gamma $, the only FLRW solutions that can be put in the metric static form (2) are the de Sitter and Lanczos universes for $\Gamma >0$, Minkowski and Milne for $\Gamma =0$, and anti-de Sitter for $\Gamma <0$~\cite{ref-journal14, ref-journal15}. All the solutions to the scale
factor can be obtained from (23). The resulting metric solutions, given by (3), are summarized in the following Table 
 \ref{tab1}.%

\begin{table}[H]
\caption{FLRW solutions expressible in static form.\label{tab1}}
	
\setlength{\cellWidtha}{\textwidth/3-2\tabcolsep-0.5in}
\setlength{\cellWidthb}{\textwidth/3-2\tabcolsep-0.5in}
\setlength{\cellWidthc}{\textwidth/3-2\tabcolsep+1in}
\scalebox{1}[1]{\begin{tabularx}{\textwidth}{>{\centering\arraybackslash}m{\cellWidtha}>{\centering\arraybackslash}m{\cellWidthb}>{\centering\arraybackslash}m{\cellWidthc}}
			\toprule
	\textbf{Density} \boldmath{$\Gamma$} & \textbf{Curvature} & \textbf{Metric} \\ \midrule
$\Gamma =0$ & $k=0$ & $ds^{2}=-d\tau ^{2}+d\rho ^{2}+\rho ^{2}d\Omega ^{2}$
\\ 
&  &   {Minkowski} \\ 
$\Gamma =0$ & $k=-1$ & $ds^{2}=-d\tau ^{2}+\tau ^{2}\left( \frac{d\rho ^{2}}{%
1+\rho ^{2}}+\rho ^{2}d\Omega ^{2}\right) $ \\ 
&  &   {Milne universe} \\ 
$\Gamma >0$ & $k=0$ & $ds^{2}=-d\tau ^{2}+e^{2\sqrt{\Gamma }T}\left( d\rho
^{2}+\rho ^{2}d\Omega ^{2}\right) $ \\ 
&  &   {de Sitter} \\ 
$\Gamma >0$ & $k=1$ & $ds^{2}=-d\tau ^{2}+\frac{\cosh ^{2}\left( \sqrt{%
\Gamma }T\right) }{\Gamma }\left( \frac{d\rho ^{2}}{1-\rho ^{2}}+\rho
^{2}d\Omega ^{2}\right) $ \\ 
&  &   {Lanczos (1)} \\ 
$\Gamma >0$ & $k=-1$ & $ds^{2}=-d\tau ^{2}+\frac{\sinh ^{2}\left( \sqrt{%
\Gamma }T\right) }{\Gamma }\left( \frac{d\rho ^{2}}{1+\rho ^{2}}+\rho
^{2}d\Omega ^{2}\right) $ \\ 
&  &   {Lanczos (2)} \\ 
$\Gamma <0$ & $k=-1$ & $ds^{2}=-d\tau ^{2}+\frac{\sin ^{2}\left( \sqrt{\left
\vert \Gamma \right \vert }T\right) }{\left \vert \Gamma \right \vert }%
\left( \frac{d\rho ^{2}}{1+\rho ^{2}}+\rho ^{2}d\Omega ^{2}\right) $ \\ 
&  &   {Anti-de Sitter} \\ 
			\bottomrule
		\end{tabularx}}
\end{table}

%

It is valuable to notice the amount of symmetry induced by the choice of $N=1$ and $b=b(\tau )$. In the initial static frame given by (2), $\mathcal{L}_{k}g_{\mu \nu }=0$ is satisfied for both $K_{t}=\partial _{t}$ and $K_{\phi }=\partial _{\phi }$. Since spatial rotations induce the existence of two others, we have at least four Killing vectors in all the spacetimes that our transformations can include. Furthermore, since for an n-dimensional maximal symmetric space, we have $n(n+1)/2$ Killing vectors, FLRW solutions, in general, have a minimum of six Killing vectors, indicating the constancy of three-dimensional spatial curvature. For the cases summarized in Table \ref{tab1}, we can see that the four-dimensional scalar curvature is equal to $R=12\Gamma $, ensuring the identification of the integration constant in (23) and (24) as the constant curvature in the Riemann tensor $R_{\mu \nu \alpha \beta }=\Gamma (g_{\mu \alpha }g_{\nu \beta }-g_{\mu \beta }g_{\nu \alpha })$. Thus, each of the cosmological solutions in this section implies maximal symmetry, with a total of ten Killing vectors~\cite{ref-book5}.

\section{From Static Coordinates ($\textbf{t,r}$) to Static Coordinates ($\mathbold{\tau ,\rho}$)}\label{sec4}

This assumption includes many subcases, of which Minkowski,
Schwarzschild, and de Sitter spacetimes can be seen as archetypical. Having static coordinates is the same as having $\dot{N}=\dot{b}=0$. Then, only the first row survives in (17), which leads to%
\begin{equation}
b^{\prime }+b^{\prime \prime }\rho -\frac{b^{\prime 2}}{b}\rho =\left( \frac{%
N^{\prime }}{N}+\frac{g^{\prime }}{g}\right) \left( b+b^{\prime }\rho
\right) .  \tag{25}
\end{equation}%
This can be rewritten as%
\begin{equation}
\frac{d\left( b+b^{\prime }\rho \right) }{d\rho }-\left( b+b^{\prime }\rho
\right) \frac{b^{\prime }}{b}=\left( \frac{N^{\prime }}{N}+\frac{g^{\prime }%
}{g}\right) \left( b+b^{\prime }\rho \right) .  \tag{26}
\end{equation}%
Dividing by the factor $b+b^{\prime }\rho =dr/d\rho $, this is equivalent to 
$d\ln \left( b+b^{\prime }\rho \right) =d\ln Ngb$. Integration yields%
\begin{equation}
b+b^{\prime }\rho =\frac{dr}{d\rho }=2\alpha Nbg,  \tag{27}
\end{equation}%
where $\alpha $ is a constant. Remember that, in this section, $bg$ only
depends on $\rho $, which means that one could obtain the explicit form for $r=r(\rho )$ directly by integration. Furthermore, consider that, from (10), the function $f$ is given by%
\begin{equation}
f=4\alpha ^{2}N^{2}.  \tag{28}
\end{equation}

Since $r$ is a function of $\rho $ and vice versa, the time-like variables $\tau $ and $t$ are the same except for a multiplicative constant. As a check of consistency, take the trivial possibility $N=bg=1$. Then, by selecting $\alpha =1/2$, one obtains $f=1$. Therefore, both (18) and (2) will be the same Minkowski spacetime.

The last two relations are of paramount importance for the remainder of this
section.

\subsection{Near Horizon Coordinates}

 We take a further step beyond Minkowski, and take, as a simple possibility, $N=\rho $, preserving $bg=1$. Now, (27) is the same as $d\left( b\rho \right)/d\rho =2\alpha \rho $, with solution $b=\alpha \rho +\beta \rho ^{-1}$.
That is, from $r=b\rho $, we obtain the relation%
\begin{equation}
r=\alpha \rho ^{2}+\beta .  \tag{29}
\end{equation}%
From it, $\alpha N^{2}=r-\beta $, and (28) is%
\begin{equation}
f=4\alpha \left( r-\beta \right) .  \tag{30}
\end{equation}%
What is the meaning of these results? Clearly, the chosen relations
correspond to the Rindler (1+1) metric $ds^{2}=-\rho ^{2}d\tau ^{2}+d\rho ^{2}$ in (18) when considering constant angles. However, even this subdimensional identification cannot globally represent Rindler space. In fact, we have been using $r=b\rho $ all along, and it comes from the angular terms.
Furthermore, we contemplate the substitution of (30) into (2): it prohibits transformation to Minkowski, contrary to the Rindler case.

The coordinate system induced by $N=\rho $ can be obtained by making the identification $\alpha =(8M)^{-1}$ and $\beta =2M$. It turns out that for the Schwarzschild case, $\rho $ corresponds to a radial coordinate that measures the proper distance at certain $r$ very near the event horizon $r=2M $~\cite{ref-book6}. A simple derivation of near horizon coordinates is performed in Appendix \ref{app1}. In particular, compare (29) with Equation~(A5).

\subsection{Tortoise Coordinates}

  Assume conformal flatness at constant angles. That is, we impose $N=bg$, and then (18) at $d\Omega ^{2}=0$ reduces to%
\begin{equation}
ds_{(2)}^{2}=N^{2}(-d\tau ^{2}+d\rho ^{2}).  \tag{31}
\end{equation}%
while (27) is now%
\begin{equation}
dr=2\alpha N^{2}d\rho .  \tag{32}
\end{equation}

For a moment, consider what occurs for a nontrivial case, e.g., let us take $N^{2}=\rho $. Then the solution to (31) is $r=b\rho =\alpha \rho ^{2}+\beta$. That is, one obtains the same dependence for $r$ seen in (29). However, in that case, we imposed $bg=1$ with $N^{2}=\rho ^{2}$, while in the actual case, we have $b^{2}g^{2}=N^{2}=\rho $. Then, (28) leads to $f=4\alpha^{2}\rho =4\sqrt{\alpha ^{3}\left( r-\beta \right) }$, with a very different behaviour than the local Rindler radial form. Less simple forms appear by considering other values for $k$ in $N=\rho ^{k}$.

However, in order to obtain more realistic metrics, another point of view is more useful here. As we emphasized before, for $\dot{N}=\dot{b}=0$, we have that $r$ is a function of $\rho $. We assume invertibility in a suitable patch, such as the exterior to the event horizon in black holes.
This allows us to consider $N=N(r)$, and by using (28) in (32), we have%
\begin{equation}
d\rho =2\alpha f^{-1}dr.  \tag{33}
\end{equation}%
We are allowed to select $2\alpha =1$, and the result is the known
Regge--Wheeler coordinate, for any spacetime characterized in the metric form given by (2). An instance is the Schwarzschild case $f=1-2M/r$, where $2M$ is the Schwarzschild radius. Integration leads to $\rho ^{\ast }=r+2M\ln \left[ r/\left( 2M\right) -1\right] $ for the exterior solution. This is the known Tortoise coordinate, which pushes the horizon event $r=2M$ to $-\infty$ in the radial coordinate $\rho ^{\ast }$. Although it does not allow for a continuous crossing of the event horizon, it is useful to obtain the Eddington--Finkelstein and Kruskal--Szekeres coordinates, via some exponentiations and rotations~\cite{ref-book4, ref-book5}.

Furthermore, take into account that one needs $r=r(\rho )$ in order to obtain $N=N(\rho )$ in (31). For the Schwarzschild spacetime, the mentioned transformation to Tortoise yields, after exponentiation, $\exp \left( \rho
^{\ast }/(2M)-1\right) =\left( r/(2M)-1\right) \exp \left( r/(2M)-1\right) $%
. Inversion involves the Lambert W function in the following form:%
\begin{equation}
r=2M\left \{ 1+W\left[ \exp \left( \frac{\rho ^{\ast }}{2M}-1\right) \right]
\right \} ,  \tag{34}
\end{equation}%
from which it is clear that the turning point from positive to negative $\rho
^{\ast }$ is approximately $1.278$ times the Schwarzschild radius. Furthermore, $%
\rho ^{\ast }\rightarrow -\infty $ when $r\rightarrow 2M$.

In Ref.~\cite{ref-journal16}, this line of thinking about the (1 + 1) conformal property is generalized. The result is the analysis of several possibilities, which includes the Kruskal--Szekeres transformations among another proposals not considered before. However, to obtain them, one must abandon the assumption that $N=N(\rho )$: it is more natural to consider an inverse route to the one explored in this article, considering the inverse transformation of (6). 

\subsection{Isotropic Coordinates}

  This system maintains explicit spherical symmetry for the spacetimes considered while putting the metric in a conformal flat form when time is constant in (2). This is achieved by transforming (2) into $-fdt^{2}+\lambda^{2}\left( d\rho ^{2}+\rho ^{2}d\Omega ^{2}\right) $. Here, $f$ is obtained as a function of $\rho $, and $\lambda ^{2}$ is the conformal factor for a flat three-dimensional space at a constant $t$. For the Schwarzschild spacetime, for instance, isotropic coordinates allow us to match the spacetime directly with the weak Newtonian metric.

Note that we just need to make the association $f\propto N^{2}$ in (18), which is consistent with relation (28), as well as with $\lambda =b$ when $g=1$ is set. Notice the distinction with case (a): now we have $N\neq 1$ in general.

The two relations (27) and (28) imply that

\begin{equation}
\frac{d\rho }{\rho }=\frac{dr}{r\sqrt{f}}.  \tag{35}
\end{equation}%
Here, we also use the fact that $r=b\rho $ and choose adequate signs
compared with the general Tortoise case (33). Here, it is also convenient to have the explicit form $f=f(r)$ in order to obtain the relation between $\rho$ and $r$. Furthermore, again, we take the Schwarzschild case in order to test that the developments yield the known solution. As before, we select $f=1-2M/r$, which allows us to express (35) in the following form:%
\begin{equation}
\frac{d\rho }{\rho }=\frac{dr}{M\sqrt{\left( \frac{r}{M}-1\right) ^{2}-1}}. 
\tag{36}
\end{equation}%
The solution is $\ln (\beta \rho )=\cosh ^{-1}\left( r/M-1\right) $, where $\beta$ is a positive constant. Performing little algebraic manipulations, we obtain%
\begin{equation}
\frac{\beta ^{2}\rho ^{2}+1}{\beta \rho }=2\left( \frac{r}{M}-1\right) . 
\tag{37}
\end{equation}%
Here, we ensure that the asymptotic behaviors of the radial coordinates $\rho $ and $r$ are the same as they approach infinity, that is, we impose that $\rho
\rightarrow r$ for $r\gg 2M$. This leads to $\beta =2/M$ in (37). Solving for $r$, we have%
\begin{equation}
r=\rho \left( 1+\frac{M}{2\rho }\right) ^{2}.  \tag{38}
\end{equation}%
Substituting this in both $f=1-2M/r$ and $b=r/\rho $, and also by recalling that $N^{2}=f$ and $g=1$, we have that (18) can be rewritten as

\begin{equation}
ds^{2}=-\left( \frac{1-\frac{M}{2\rho }}{1+\frac{M}{2\rho }}\right)
^{2}d\tau ^{2}+\left( 1+\frac{M}{2\rho }\right) ^{4}\left( d\rho ^{2}+\rho
^{2}d\Omega ^{2}\right) .  \tag{39}
\end{equation}%
This is the usual isotropic form for the Schwarzschild spacetime~\cite{ref-book3}. 

\subsection{Lema\^{\i}tre-Type for Schwarzschild}

  Lema\^{\i}tre coordinates $(\tau ,\rho )$ for the Schwarzschild spacetime are such that the metric takes the form%
\begin{equation}
ds^{2}=-N^{2}d\tau ^{2}+\frac{\kappa }{r}d\rho ^{2}+r^{2}d\Omega ^{2}, 
\tag{40}
\end{equation}%
with $N=1$ and $\kappa =2M$~\cite{ref-journal1, ref-book4}. By considering (28), this form implies that $2\alpha =1$ if one takes $t=\tau $, and the fact that $f=1$ in (2) indicates that we are constrained to Minkowski spacetime from the very start. We can just take into account that $rdr^{2}=\kappa
d\rho ^{2}$ to show that $4r^{3}=9\kappa \rho ^{2}$ is an appropriate
transformation for this case.

All the conditions imposed in this section, in particular that $N$ and $b$ depend only on $\rho $, need to be relaxed in order to obtain the Lema\^{\i}tre form for general spacetimes. However, for the conditions in this very section, we can obtain a form that resembles it for the Schwarzschild spacetime when $t$ is constant.

Consistence with (40) demands that $b^{2}g^{2}=2M/r$, and from (28), we have $4\alpha ^{2}N^{2}=1-2M/r$. Then (27) can be expressed as $rdr/\sqrt{(r/2M)-1}=2Md\rho $, which is solved by%
\begin{equation}
\rho =\frac{4M}{3}\left( \frac{r}{2M}+2\right) \sqrt{\frac{r}{2M}-1}. 
\tag{41}
\end{equation}%
We set an integration constant to zero. It maintains the range of $\rho $ from $0$ to $\infty $ when $r$ goes from $2M$ to $\infty $, which is the exterior patch of the black hole. As a matter of contrast, remember that the range is pushed from $(2M,\infty )$ in $r$ to $(-\infty ,\infty )$ in $\rho ^{\ast }$, the Tortoise coordinate appearing in (34). Notice also that for $r\gg 2M$, the behavior is $4r^{3}=9(2M)\rho ^{2}$, the same for the Minkowski case mentioned above. That is, we have explicit asymptotic flatness for the Schwarzschild spacetime. However, as interesting as it may sound, the range of coordinates induced by the transformation (41) limits its validity to the exterior of the event horizon in the Schwarzschild spacetime. By contrast, the most attractive feature of Lema\^{\i}tre coordinates is precisely that they allow for a continuous crossing of the event horizon~\cite{ref-journal1, ref-book4}.

\section{Other Relevant Cases}\label{sec5}

There are several cases of interest in which it is necessary to
abandon the assumption that $N=N(\rho )$ and $b=b(\rho )$. In particular, in the Kruskal--Szekeres coordinates, $N$ depends on both $\tau $ and $\rho $. The same is true for the factor $bg$ in Lema\^{\i}tre coordinates for Schwarzschild. As we mentioned before, the first case readily appears by taking the inverse transformation, instead of the one used in this article.
That is, going from the metric form of (3) to that of (2), as in~\cite{ref-journal13,ref-journal16}, one can obtain those coordinates within our scheme, albeit indirectly. 

\subsection{Kruskal--Szekeres (Indirect Route)}

  Maintain $\dot{N}=\dot{b}=0$ in such a way that the relations in the first
part of Section \ref{sec4} are still valid. Take the case of $N=bg\rho $, which converts (18) to%
\begin{equation}
ds_{(2)}^{2}=n^{2}(-\rho ^{2}d\tau ^{2}+d\rho ^{2})+r^{2}d\Omega ^{2}, 
\tag{42}
\end{equation}%
where we define $n=N/\rho $. Now, we, again, have the conformal Rindler form at hypersurfaces where $d\Omega ^{2}=0$.

Notice that now (27) and (28) imply the relation%
\begin{equation}
\frac{d\rho }{\rho }=2\alpha \frac{dr}{f}.  \tag{43}
\end{equation}%
Its solution is $\rho =ke^{\rho ^{\ast }}$, where $\rho ^{\ast }$ is the generic solution to (33). Actually, this exponentiation constitutes the starting point of the textbook route for obtaining Kruskal--Szekeres -via the intermediate definition of Eddington--Finkelstein coordinates~\cite{ref-book2, ref-book4, ref-book5}.

Let us define $U=\rho e^{\tau }$, $V=-\rho e^{-\tau }$, that allows us to
express the metric (42) as $-n^{2}dUdV+r^{2}d\Omega ^{2}$. Now, we obtain a suggestive representation, by performing the custom rotation $V=T-X$ and $U=T+X$, in such a way that we have%
\begin{equation}
ds_{(2)}^{2}=n^{2}(-dT^{2}+dX^{2})+r^{2}d\Omega ^{2}.  \tag{44}
\end{equation}%
It is useful to compare this relation with (31). There, $N=N(\rho )$ in the $(\tau ,\rho ,\theta ,\phi )$ system, while in (44), we have $n=n(T,X)$. Even more, in the Kruskal--Szekeres case, it is better to keep the conformal factor in terms of the original radial coordinate $r$.

More precisely, for the Schwarzschild spacetime, the association $n^{2}\rho
^{2}=8M(1-2M/r)$ can be made. Furthermore, the solution to (43) is $\rho
=ke^{\rho ^{\ast }}$, where we have the (rescaled-) Tortoise coordinate, given by%
\begin{equation}
4M\rho ^{\ast }=r+2M\ln \left( \frac{r}{2M}-1\right) .  \tag{45}
\end{equation}%
Here, we choose $2\alpha =1/(4M)$ in (43). Furthermore, with $k\sqrt{2M}=1$ in $\rho =ke^{\rho ^{\ast }}$, we have%
\begin{equation}
\rho ^{2}=\frac{1}{2M}\left( \frac{r}{2M}-1\right) e^{\frac{r}{2M}}. 
\tag{46}
\end{equation}%
Substituting it in $n^{2}\rho ^{2}=4r_{s}(1-r_{s}/r)$, (44) can be written as%
\begin{equation}
ds_{(2)}^{2}=\frac{32M^{3}}{r}e^{-\frac{r}{2M}}(-dT^{2}+dX^{2})+r^{2}d\Omega
^{2},  \tag{47}
\end{equation}%
which is the usual Kruskal--Szekeres form for the Schwarzschild spacetime.

\subsection{Lema\^{\i}tre Coordinates}

 The Lema\^{\i}tre form for the Schwarzschild spacetime appears by imposing $N=1$ and $b^{2}g^{2}=2M/r$ in (18)~\cite{ref-journal1, ref-book4}. Compared with the last example in Section \ref{sec4}, here, we consider $r=r(\tau ,\rho )$.

{Remembering that $r=b\rho $, we have $b+b^{\prime }\rho =\partial r/\partial
\rho $ and $\dot{b}\rho =\partial r/\partial \tau $. Then, Equation~(10)} can be written as%
\begin{equation}
1-\frac{2M}{r}=\frac{r}{2M}\left( \frac{\partial r}{\partial \rho }\right)
^{2}-\left( \frac{\partial r}{\partial \tau }\right) ^{2}.  \tag{48}
\end{equation}%
This form suggests to propose $r=A(\rho -\tau )^{n}$, as it allows us to
factorize. Substitution and simplification yield%
\begin{equation}
\left[ nA\left( \rho -\tau \right) ^{n-1}\right] ^{2}=\frac{2M}{r}.  \tag{49}
\end{equation}%
Again, we use $r=A(\rho -\tau )^{n}$, leading to $n^{2}A^{3}\left( \rho -\tau
\right) ^{3n-2}=2M$. This alone implies that $n=2/3$ and also that $\left(
2/3\right) ^{2}A^{3}=2M$. That is, $r=r(\tau ,\rho )$ is solved by%
\begin{equation}
r=\left[ \frac{9}{2}M(\rho -\tau )^{2}\right] ^{\frac{1}{3}}.  \tag{50}
\end{equation}

Now, consider that, for this case, (11) and (12) can be recast as

\begin{equation}
\frac{\partial t}{\partial \tau }=\frac{1}{1-\frac{2M}{r}}\sqrt{\frac{r}{2M}}%
\left( \frac{\partial r}{\partial \rho }\right)   \tag{51}
\end{equation}%
and%
\begin{equation}
\frac{\partial t}{\partial \rho }=\frac{1}{1-\frac{2M}{r}}\sqrt{\frac{2M}{r}}%
\left( \frac{\partial r}{\partial \tau }\right) ,  \tag{52}
\end{equation}%
respectively. Furthermore, the differential of (50) is%
\begin{equation}
dr=\sqrt{\frac{2M}{r}}(d\rho -d\tau ),  \tag{53}
\end{equation}%
where we use (50) in the form of $(\rho -\tau )^{-\frac{1}{3}}=\left( 
\sqrt{9M/2}\right) ^{\frac{1}{3}}r^{-1/2}$. This indicates that $\partial
r/\partial \rho =-\partial r/\partial \tau =\sqrt{2M/r}$, which can be
substituted into (51) and (52). In turn, this allows us to obtain $%
dt=(\partial t/\partial \tau )d\tau +(\partial t/\partial \rho )d\rho $. The result is%
\begin{equation}
dt=\frac{1}{1-\frac{2M}{r}}\left( d\tau -\frac{2M}{r}d\rho \right) . 
\tag{54}
\end{equation}%
The inversion of (53) and (54) is%
\begin{equation}
d\tau =dt+\sqrt{\frac{2M}{r}}\frac{dr}{1-\frac{2M}{r}}  \tag{55}
\end{equation}%
and%
\begin{equation}
d\rho =dt+\sqrt{\frac{r}{2M}}\frac{dr}{1-\frac{2M}{r}}.  \tag{56}
\end{equation}%
These two relations, together with (50), constitute the usual Lema\^{\i}tre transformation~\cite{ref-journal1, ref-book4}, with metric%
\begin{equation}
ds^{2}=-d\tau ^{2}+\frac{2M}{r}d\rho ^{2}+r^{2}d\Omega ^{2}.  \tag{57}
\end{equation}%

\section{Discussion}\label{sec6}

In this article, we present a consistent and self-contained
method to obtain distinct representations of spherically symmetric
spacetimes that share the metric form%
\begin{equation*}
ds^{2}=-fdt^{2}+f^{-1}dr^{2}+r^{2}d\Omega ^{2}
\end{equation*}%
in static coordinates. We transform it to the general form%
\begin{equation*}
ds_{(2)}^{2}=-N^{2}d\tau ^{2}+b^{2}\left( g^{2}d\rho ^{2}+\rho ^{2}d\Omega
^{2}\right) ,
\end{equation*}%
where $N$ and $b$, in general, are functions of $(\tau ,\rho )$ and $g=g(\rho
) $. This constitutes a special case of ADM decomposition, as we argue in Section \ref{sec2}, and this identification motivated us to analyze FLRW solutions. As summarized at the end of Section \ref{sec3}, there are only six different FLRW solutions that satisfy the aforementioned transformation. However, more general homogeneous cosmological solutions have been analyzed in terms of the ADM formalism (see, for instance, Ref.~\cite{ref-book7} and references therein).

Section \ref{sec4} is dedicated to analyzing metrics that transform from the initial
static version to another static form, in the sense that $N$ and $b$ are
independent of $\tau $ in the new system. Many of the known analytical
solutions to Einstein's equations satisfy this requirement. In particular,
by making small assumptions, we obtain Near Horizon Coordinates, as well as
Tortoise and Isotropic coordinates, for the Schwarzschild solution. At the
end of Section \ref{sec3}, we also briefly obtain a Lema\^{\i}tre-like case on
hypersurfaces with $t$ being constant. However, it is more of an illustrative
example since it does not share desirable properties of the correct Lema%
\^{\i}tre system, such as allowing for a continuous crossing of the event
horizon, or the interpretation of the time coordinate with a free-falling
observer, that is, the same time $\tau $ as Painlev\'{e}--Gullstrand coordinates.

In the final part, we include two important examples: the
Kruskal--Szekeres and Lema\^{\i}tre coordinates. Although both are obtained
with relations from the previous sections, the two coordinate systems
needed some additional assumptions: this is the result of allowing $b=b(\tau
,\rho )$ in the transformed system.

Throughout the developments, we emphasize the Schwarzschild spacetime.
However, the method may be readily applied to other spacetimes, such as
Reissner--Nordstr\"{o}m, de Sitter, anti-de Sitter, among others~\cite{ref-journal13, ref-journal16, ref-journal17}.

It is worth highlighting the importance of the method we use, as well as the results. We see how imposing symmetries at the level of metric transformations can lead to uniquely identifying which spacetime is described. This is the case in the FLRW cases, near horizons, and isotropic coordinates, for instance. Furthermore, it is clear how less restricted variables $N$ and $b$ must be accompanied by another class of assumptions, as is the case with the Kruskal--Szekeres and Lema\^{\i}tre representations of the Schwarzschild spacetime. As we mentioned already, there is a direct relation with the ADM formulation. However, it is also instructive to notice other implications of the symmetries. The initial static form directly implies the presence of four Killing vectors. As noted at the end of Section \ref{sec3}, by imposing $N=1$ and $b=b(\tau )$ in the transformations, they must grow to a total of ten. That this, the FLRW solutions that can be included in our ansatz are the maximal symmetric ones. However, this does not happen in the general case. Just as an example, isotropic coordinates can describe the Schwarzschild spacetime, which keeps just the four initial Killing vectors. However, we could also obtain the isotropic coordinates for the de Sitter space, which has the ten mentioned Killing vectors.

Finally, Einstein's field equations are implicitly satisfied in the Schwarzschild examples. Even more interestingly, they appear---in the form of the two Friedmann equations---when $N=1$ $b=b(\tau )$. We argue that this type of symmetry manifestation could be of help to obtain exact solutions in other gravitational theories or may be used as a guiding principle to discern desirable properties of theories such as $f(R)$ gravity, higher-derivative versions, as well as in other types of modified gravity theories (see, for instance, \cite{ref-journal18, ref-journal19, ref-journal20}).      
\vspace{6pt} 

\authorcontributions{Conceptualization, E.A.L. and A.S.-R.; methodology, E.A.L.; validation, E.A.L. and A.S.-R.; formal analysis, E.A.L. and A.S.-R.; investigation, E.A.L. and A.S.-R.; writing---original draft preparation, E.A.L.; writing---review and editing, E.A.L. and A.S.-R. All authors have read and agreed to the published version of the manuscript. All authors have read and agreed to the published version of the manuscript.}
\funding{This work was partially supported by Coordinación General para el Fomento a la Investigación Cient\'{i}fica e Innovación del Estado de Sinaloa (CONF\'{I}E).}
\dataavailability{No new data were created or analyzed in this study. Data sharing is not applicable to this article.}
\acknowledgments{The authors thank J. A. Nieto and B. Mart\'{i}nez-Olivas for helpful comments. ASR acknowledges a graduate fellowship grant by CONAHCYT-M\'{e}xico. The authors also thank the anonymous referees for their comments that helped to improve the quality of this work.}

\conflictsofinterest{The authors declare no conflicts of interest.}

\appendixtitles{no} 
\appendixstart
\appendix
\section[\appendixname~\thesection]{}\label{app1}
Start by noticing that a (shell) observer fixed at a Schwarzschild
coordinate $r$ measures a radial spacelike distance ($dt=d\theta =d\phi $)
given by
\begin{equation}
ds=\frac{dr}{\sqrt{1-\frac{2M}{r}}},  
\end{equation}%
that directly yields the known radial length contraction for the
Schwarzschild observer as one approaches $r_{s}=2M$. This induces a radial
coordinate that can be rewritten as%
\begin{equation}
d\varrho =\frac{rdr}{\sqrt{r^{2}-2Mr}}. 
\end{equation}%
Defining $w=r^{2}-2Mr$, we have $dw=2(r-M)dr$; the last expression is the
same as%
\begin{equation}
d\varrho =\frac{dw}{2\sqrt{w}}+\frac{Mdr}{\sqrt{(r-M)^{2}-M^{2}}},
\end{equation}%
which, when integrated, is%
\begin{equation}
\varrho =\sqrt{r(r-2M)}+M\cosh ^{-1}\left( \frac{r}{M}-1\right) . 
\end{equation}%
Note that the integration constant is zero in such a way that $\varrho =0$
when $r=2M$.

Now, since $\cosh ^{-1}\left( 2x^{2}+1\right) =2\sinh x$ for $x\geq 0$, the
second term is $2M\sinh \sqrt{r/2M-1}$. For an $r$ very near to $r=2M$, it can
be approximated to $2M\sqrt{r/2M-1}$, while the first term is $\sqrt{2M(r-2M)%
}=2M\sqrt{r/2M-1}$. That is, when $r$ is very near to $2M$, (A4) is $%
\varrho \approx 2\sqrt{2M(r-2M)}.$ In terms of the new coordinate,%
\begin{equation}
r=\frac{\varrho ^{2}}{8M}+2M,
\end{equation}%
and by substituting it in the Schwarzschild metric (at constant angles), we
obtain%
\begin{equation}
ds^{2}=-\left( \frac{\varrho ^{2}}{\varrho ^{2}+16M^{2}}\right)
dt^{2}+d\varrho ^{2}. 
\end{equation}%
Near the horizon, $\varrho ^{2}+16M^{2}\approx 16M^{2}$, and scaling the
time coordinate as we have scaled the time coordinate $T=t/4M$, we obtain%
\begin{equation}
ds^{2}\approx -\varrho ^{2}dT^{2}+d\varrho ^{2}, 
\end{equation}%
that is, the (1 + 1) Rindler space that can be useful to make an association
between accelerated frames in the Minkowski space with the Schwarzschild space,
remarkably indicating the Unruh effect and Hawking radiation.

\begin{adjustwidth}{-\extralength}{0cm}

\reftitle{References}

\PublishersNote{}
\end{adjustwidth}
\end{document}